\documentclass{article}
\usepackage{arxiv}
\usepackage[utf8]{inputenc} % allow utf-8 input
\usepackage[T1]{fontenc}    % use 8-bit T1 fonts

\usepackage{amsmath}
\usepackage{algorithmic}
\usepackage{xspace}
\usepackage{subfigure}
\usepackage{caption}
\usepackage{graphicx}

\usepackage{makecell}
\usepackage{threeparttable}
\usepackage{multirow}
\usepackage{booktabs}
\usepackage{CJKutf8}

\usepackage[noadjust]{cite}

\def\cn#1{\begin{CJK*}{UTF8}{gbsn}\small #1\end{CJK*}}

\newcommand{\org}[1]{\textsuperscript{#1}}

\usepackage{authblk}

\title{Kilometer-Level Coupled Modeling Using 40 Million Cores: An Eight-Year Journey of Model Development}

\renewcommand{\shorttitle}{Kilometer-Level Coupled Modeling Using 40 Million Cores}

\begin{document}

\author[ ]{Xiaohui Duan\org{a,d,*}\quad Yuxuan Li\org{b,*}\quad Zhao Liu\org{b,d,*}\quad Bin Yang\org{b,d}\quad Juepeng Zheng\org{f}\quad Haohuan Fu\org{b,d,\dag} \quad
Shaoqing Zhang\org{c,e,\dag}\quad Shiming Xu\org{b,d,\dag}\quad Yang Gao\org{c,e}\quad Wei Xue\org{b,d}\quad Di Wei\org{b}\quad Xiaojing Lv\org{g} \quad
Lifeng Yan\org{a}\quad Haopeng Huang\org{b,d}\quad Haitian Lu\org{d}\quad Lingfeng Wan\org{c,e}\quad Haoran Lin\org{a}\quad Qixin Chang\org{a}\quad
Chenlin Li\org{d}\quad Quanjie He\org{d}\quad Zeyu Song\org{b}\quad Xuantong Wang\org{b}\quad Yangyang Yu\org{c,e}\quad Xilong Fan\org{a}\quad
Zhaopeng Qu\org{j}\quad Yankun Xu\org{j}\quad Xiuwen Guo\org{c,e}\quad Yunlong Fei\org{c,e}\quad Zhaoying Wang\org{c,e}\quad Mingkui Li\org{c,e} \quad
Yingjing Jiang\org{c,e}\quad Lv Lu\org{c,e}\quad Liang Su\org{j}\quad Jiayu Fu\org{b}\quad Peinan Yu\org{b}\quad Weiguo Liu\org{a,d,\dag}\quad Lixin Wu\org{c,e,\dag}\quad
Lanning Wang\org{i}\quad Xin Liu\org{h}\quad Dexun Chen\org{d}\quad Guangwen Yang\org{b,d}
}
\thanks{These authors contribute equally to this work.}
\thanks{Corresponding authors are Haohuan Fu (haohuan@tsinghua.edu.cn), Shaoqing Zhang (szhang@ouc.edu.cn), Shiming Xu (xusm@tsinghua.edu.cn), Weiguo Liu (weiguo.liu@sdu.edu.cn), and Lixin Wu(lxwu@ouc.edu.cn)}

\affil[a]{School of Software, Shandong University, China}
\affil[b]{Tsinghua University, China}
\affil[c]{Ocean University of China, China}
\affil[d]{National Supercomputing Center in Wuxi, China} 
\affil[e]{Laoshan Laboratory, China}
\affil[f]{School of Artificial Intelligence, Sun Yat-Sen University, China}
\affil[g]{China Ship Scientific Reseach Center, China}
\affil[h]{National Research Center of Parallel Computer Engineering and Technology, China}
\affil[i]{Beijing Normal University, China}
\affil[j]{Qingdao GROSCI Technology Group, China}

\maketitle

\def\fino{265\xspace}
\def\fina{340\xspace}
\def\finc{222\xspace}
\def\stc{1.79\xspace}
\def\fini{838\xspace}
\def\finl{5050\xspace}
\def\ncore{39,702,000\xspace}

\begin{abstract}
With current and future leading systems adopting heterogeneous architectures, adapting existing models for heterogeneous supercomputers is of urgent need for improving model resolution and reducing modeling uncertainty. This paper presents our three-week effort on porting a complex earth system model, CESM 2.2, to a 40-million-core Sunway supercomputer. Taking a non-intrusive approach that tries to minimizes manual code modifications, our project tries to achieve both improvement of performance and consistency of the model code. By using a hierarchical grid system and an OpenMP-based offloading toolkit, our porting and parallelization effort covers over 80\% of the code, and achieves a simulation speed of \fina SDPD (simulated days per day) for 5-km atmosphere, \fino SDPD for 3-km ocean, and \finc SDPD for a coupled model, thus making multi-year or even multi-decadal experiments at such high resolution possible.
\end{abstract}


%%%%%%%%%%%%%%%%%%%%%%%%%%%%%%%%%%%%%%%%%%%%%%%%%%%%%%%%%%%%%%
\section{Justification for the Best Application Award (Climate Modeling)}

We form a non-intrusive yet efficient workflow to port CESM 2.2 to a 40-million-core heterogeneous supercomputer, in around three weeks. Maintaining the consistency of the code, we improve from simulating \stc days to \finc days per day (enabling multi-year or even multi-decadal ultra-high-resolution climate modeling).

\section{Performance Attributes}
\begin{table}[htb]
\caption{Performance Attributes of our work}\label{tab:perfattr}
\centering
\begin{tabular}{ll}
\hline
\textbf{Attribute Title}       & \textbf{Attribute Value}                                                                                                                                        \\ \hline
\textbf{Category of achievement}                  & \makecell[l]{Scalability; \\ time-to-solution}                                                                               \\ \hline
\textbf{Type of method used}                  & \makecell[l]{Finite element with \\ implicit/explicit\\time-stepping}                                                                               \\ \hline

\textbf{\makecell[l]{Results reported on the \\basis of}}    & \makecell[l]{Whole application  excluding \\ I/O and initialization}  \\ \hline
\textbf{Precision reported}           & Double precision          \\ \hline
\textbf{System scale}                 & \makecell[l]{Results measured on \\full-system scale  }    \\ \hline
\textbf{Measurement mechanism}        & Timers         \\ \hline
\textbf{Performance Result}           & \makecell[l]{CAM \fina SDPD \\POP \fino SDPD\\CICE \fini SDPD\\LND \finl SDPD \\ Coupled \finc SDPD \\  }    \\ \hline
\end{tabular}
\end{table}
%%%%%%%%%%%%%%%%%%%%%%%%%%%%%%%%%%%%%%%%%%%%%%%%%%%%%%%%%%%
Shown in TABLE \ref{tab:perfattr}.

\section{Overview of the Problem}

\begin{figure}[ht]                                    
    \centering 
    \includegraphics[width=0.48\textwidth]{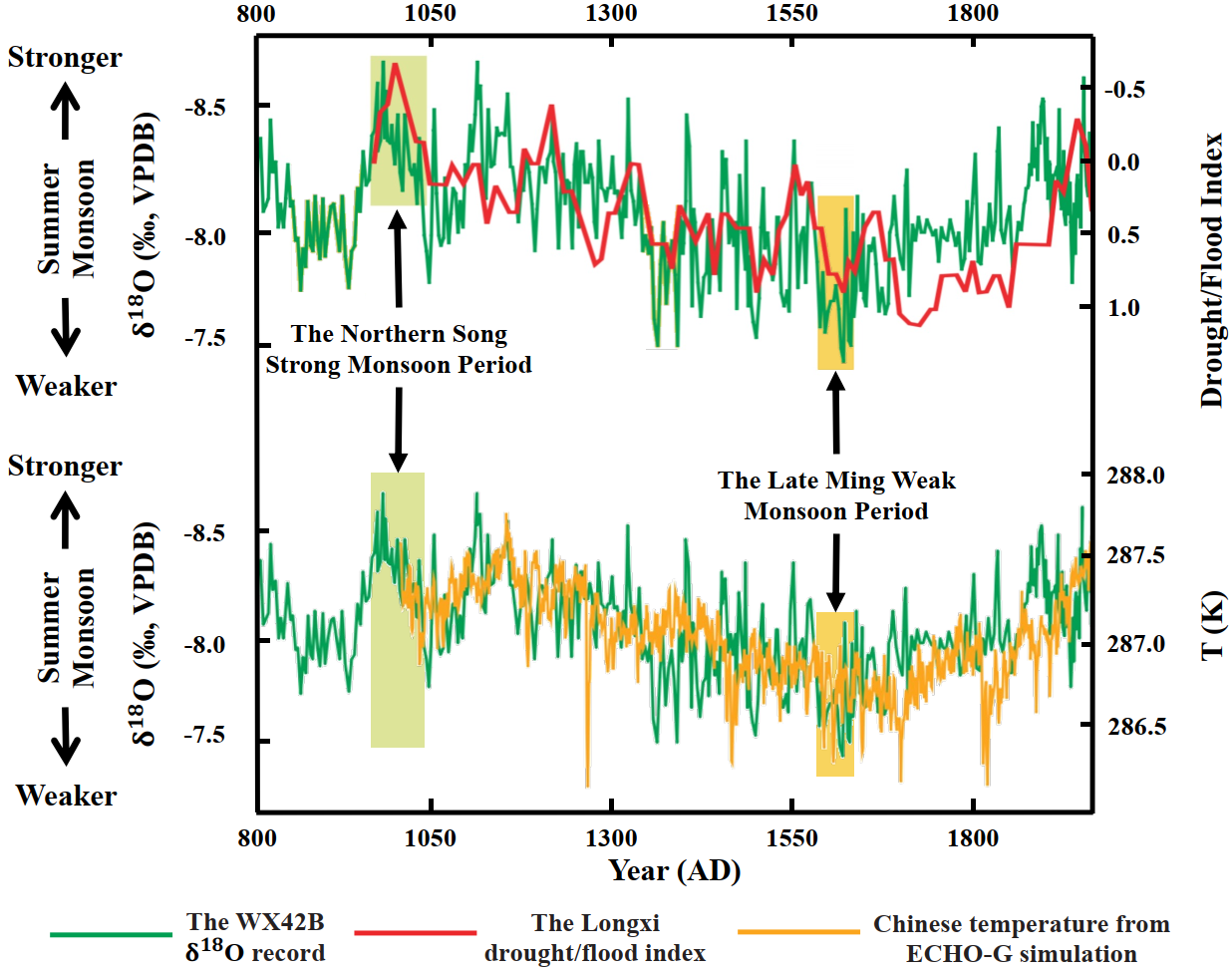}
    \caption{The drought/flood index and reconstructed temperature change since 800 A.D. (results from existing works \cite{zhang2008test-monsoon,yang2002general-ChinaTempReconstruction}), corresponding to some of the major events in history.}
    \label{fig:temp-history}   
\end{figure}

Climate, in the long history of humankind, is always considered as an important factor for the life of every single person, and the prosperity of a city, a nation, and even the entire human community on earth. Its rich variety of patterns, and unpredictable dynamics, play important roles in the rising and falling of dynasties (mostly due to the competition between the agriculture people and the nomadic tribes powers through the cycle of warm and cold climates). Figure \ref{fig:temp-history} shows the reconstructed climate change over the last few thousand years, demonstrating a precise alignments with some of the most famous ``golden" decades in human history, as well as those most chaotic times dominated by warlords. We combine the temperature reconstruction study in \cite{yang2002general-ChinaTempReconstruction} and the Asian Monsoon study in \cite{zhang2008test-monsoon} (while we generally correlate the activities of Asian Monsoon to the precipitation in eastern China, the major economic zone of ancient Chinese dynasties, there can be uncertainty issues in decadal scales \cite{zhang2010linking}) for the last two millennia. A most prominent alignment of warm temperature and abundant precipitation was around the year of 1000 A.D., the early years of the Northern Song Dynasty, one of the most economic-active period of time in China's ancient history. In contrast, the decrease of both temperature and Asian Monsoon strength correlated with the frequent uprising of peasants and the collapse of the Ming Dynasty.

The dynamic and complex feature also makes the climate a most challenging and most intriguing topic for scholars since ancient times. Kongming (\cn{孔明}), the best established icon of wisdom in Chinese literature \cite{guanzhong2018romance}, made his most important moment for magically predicting a sudden change from the normal northwest wind in the winter to the southeast wind, which enabled the victory of the Sun-Liu Union from the south over Cao's army from the north, through a big fire at the Red Cliff. While we can not ascertain such prediction capabilities in ancient times, the literature demonstrates our obsession in understanding and predicting the climate system.

While such predictions in the old days mostly rely on observations and pragmatic judgements, modern climate services rely heavily on computer-based modeling of the climate system. Dated back to the very first numerical weather program (NWP) on the very first electronic computer ENIAC \cite{platzman1979eniac}, the climate modelers have experienced a co-evolution between generations of supercomputers (Cray vector machines in the 1970s, IBM and Intel clusters around 2000, and the various heterogeneous supercomputers in the recent decade), and generations of climate models (with more an more components integrated, and resolutions evolving from a few hundred kilometers to a few kilometers).

We can roughly categorize the co-evolution of supercomputers and climate models into three major stages. 

The first stage is the era of climate models on vector processors (CDC 6600, Cray I, Cray X-MP4 etc.) from 1960s to 1990s \cite{mcguffie2001forty}. With architecture innovations emerging from each new generation of hardware (such as pipelines and vector units proposed by Seymour Cray), developers had to tune their model carefully to the underlying vector processors and handle issues such as vectorization and parallelization for conditional statements in physics schemes \cite{dickinson1982optimizing}.  

The second stage (1990s to 2010s) is marked by the transition from a few large but strong vector processors to numerous small but fast-connected processors (a vivid description is Cray's famous quote on comparing two oxen and 1,024 chickens \cite{mitchell1990genius}). Early successful examples include the development of a parallel global circulation model (GCM) \cite{chervin1988relationship} and an eddy-resolving ocean model \cite{chervin1990ocean,semtner1992ocean}. After the year of 2000, we start to see a spectral global circulation model that can utilize 5,120 processors of the Earth Simulator (for T1279 16-km resolution) \cite{shingu200226}, and a WRF nature run that scaled to 65,536 processors of the IBM BlueGene/L machine (5-km resolution) \cite{michalakes2007wrf}. 

For the third stage (the recent decade), climate models start to adapt to heterogeneous supercomputers. Examples include Community Atmosphere Model (CAM) on Sunway TaihuLight \cite{fu2016refactoring,fu2017redesigning}, the Consortium for Small-Scale Modeling (COSMO) on Piz Daint, and the E3SM model project for the US exascale supercomputers \cite{leung2020introduction-e3sm}. 

Recently, the installment of exascale supercomputers in different regions, has called for an aggressive step towards kilometer level climate modeling, expected to reduce the uncertainty of current climate models to a new level \cite{schar2020kilometer}. While great advantages are expected for climate models, such a step involves challenges at least from three different aspects:

\begin{itemize}
\item The need for new numerical schemes (grids and solvers), and the adaptation of the physics schemes, to accommodate the kilometer-level high resolution scenarios.
\item The demand to port, parallelize, and optimize the dynamic and physics components, including both a heavy legacy code and the aforementioned newly developed code, on the heterogeneous accelerators with a completely different compute and memory granularity from previous homogeneous processors.
\item The significantly increased complexity related to MPI communication, parallel I/O, resilience and reproducibility, at such a ultra-high-resolution and parallel scale. 
\end{itemize}

Facing such severe challenges from both the evolving scientific demands of the application side and the evolving programming demands of the architecture side, one straightforward approach might be to develop something new from scratch (such as the E3SM project that started around five years before the deployment of the US exascale system). However, the climate model itself is not only a program to simulate the climate behavior of the earth system, but rather a software form of community knowledge integration, succession, and gradual evolution. For example, various versions of the community earth system model (CESM), and its predecessor community climate system model (CCSM), have accumulated, verified, and improved international scientists' understanding of the climate change mechanism for the last few decades. 

Therefore, when migrating such a system software into a new supercomputer platform, the performance improvement and the consistency of the software description of the science, are both important factors that we need to consider. Instead of making significant code changes (as in our previous refactoring and redesigning efforts \cite{fu2016refactoring,fu2017redesigning}), we decide to take a non-intrusive approach that relies on behind-the-scene schemes and tools to facilitate a more fluent transition to a new supercomputer system. From the previous experience, this also seems a useful strategy to encourage the use of such migrated software in scientific experiments and explorations \cite{lin2020community,chang2020unprecedented,zhang2023toward}. Our major strategies are as follows:
 
(1) We adopt a hierarchical system of grids with different spatial resolutions, and a new tripolar grid generation method based on conformal mapping to handle the irregular shapes in such cases. The grid system provides a basis for the domain decomposition and parallelization afterwards, with no need to modify the numerical codes running on them. 

(2) Based on the current software ecosystem of the Sunway Supercomputer, we develop a prototype OpenMP offloading framework on the new Sunway Supercomputer. Instead of the fine-grained parallelization and memory related tuning efforts needed before, we can enable efficient utilization of heterogeneous cores for almost all compute parts of the model in an OpenMP-oriented fashion.

(3) For an unprecedented scale of around 600,000 MPI processes, we redesign the communication patterns at the initialization stage as a hierarchical tree, which successfully cuts the initialization time by over 3 times.

(4) Lastly and most importantly, we form a complete tool chain to support the migration of such a complex earth system model to the 40-million-core Sunway supercomputer. With tools covering profiling, hardware fault detection, bit-accurate result validation, and binary static call map analysis functions, we manage to cover over 80\% of the entire model, and achieve considerable speedup for entire component models. 

The current version of a 5km-atmosphere and 3km-ocean coupled model can scale to over 101,800 nodes (\ncore cores), and achieve a simulation speed of \fina SDPD for atmosphere, \fino SDPD for ocean, \fini SDPD for sea ice, and \finc SDPD for the coupled model. We think such a performance would provide a useful tool to enable scientists to look at the possibilities of generating and analyzing multi-decadal or even multi-century simulation results (similar to the 750-year we have produced before for a 25km-atmosphere and 10-km-ocean configuration \cite{chang2020unprecedented}).

%%%%%%%%%%%%%%%%%%%%%%%%%%%%%%%%%%%%%%%%%%%%%%%%%%%%%%%%%%%%%%
\section{Current State of the Art}

\subsection{Recent Efforts Towards Kilometer-Level Climate Modeling}

\begin{table*}[t]
\caption{The comparisons among recent kilometer-resolution climate modeling efforts.}
\begin{center}
\resizebox{1\linewidth}{!}{
\begin{tabular}{cccccc}
\hline
\textbf{Model} & \textbf{COSMO} & \textbf{HOMMEXX-NH } & \textbf{NICAM } & \textbf{ICON } & \textbf{CESM-HR (our work)} \\ \hline%\hline
Machine & Piz Daint & Summit & Fugaku & Levant & Sunway\\
\hline
Model & Regional Atmosphere &  Atmosphere Model  & Atmosphere & Coupled ESM  & Coupled ESM  \\
Description & Model & Dycore & Model & with Reduced Physics  & with Complete Physics \\
\hline
Domain size & Near-global & Global & Global & Global & Global \\
\hline
\multirow{2}*{Resolution} & \multirow{2}*{0.93 km} & \multirow{2}*{3 km} & \multirow{2}*{3.5 km} & 5-km ATM  & 5-km ATM  \\
&&&&5-km OCN&3-km OCN\\
\multirow{2}*{Porting process} & Fortran$\rightarrow$C++  & Fortran$\rightarrow$C++ & multi-year  & OpenACC & OpenMP offloading\\
& STELLA DSL & Kokkos library & Codesign & directives & and other tools\\
\hline
\multirow{2}*{Scale} & 4,888 GPUs  & 276,000 GPUs & 131,072 nodes & 600 AMD nodes & 101,800 processors\\
& (Tesla P100) &  (Tesla V100) &  (6,291,456 cores) &  (76,800 cores) & (\ncore cores) \\
\hline
%Performance & \\
\multirow{2}{*}{SDPD} & \multirow{2}*{15}  & \multirow{2}*{354} & 2.5  & \multirow{2}*{126} & \multirow{2}{*}{\finc}  \\
&&&(estimated from the DA flow)&&\\%(\fina/\fino/\fini for ATM/OCN/ICE)\\

\hline
\end{tabular}
}
\label{tab:comparison}
\end{center}
\end{table*}

With the boost from the successful Earth Simulator supercomputer, following the work of a 16-km atmospheric simulation \cite{shingu200226}, Satoh \emph{et. al} applied the Nonhydrostatic Iscosahedral Atmospheric Model (NICAM) to achieve 3.5-km and 7-km global simulations \cite{satoh2008nonhydrostatic} based on the Earth Simulator, and 870-m global resolution \cite{miyamoto2013deep} on K computer. In US, a highly-scalable CAM-SE dynamic core was developed to support 12.5-km resolution with a simulation speed of around 4.6 SYPD across over 172,800 CPU cores on Jaguar \cite{dennis2012cam}. The Weather Research and Forecasting (WRF) model provides a single-precision performance of 285Tflops on 437,760 cores of Blue Waters \cite{johnsen2013petascale}. With the potential of performing ultra-high resolution climate simulations demonstrated, these are still more pioneering works to try multi-day or multi-week experiments, or to explore the high-performance solvers and tuning techniques on supercomputers. Practical climate simulation at kilometer level is still out of reach.

Only with the exascale systems coming into concrete plans, we start to see ambitious projects that would eventually lead to multiple-year kilometer-scale climate simulation.

The one that started the earliest is probably the E3SM (Energy Exascale Earth System Model) project (phase 1 started in 2014) \cite{leung2020introduction-e3sm}. With a clear goal to integrate model development with the leading‐edge computational advances, E3SM targets ultra-high-resolution, and branches from the most widely used CESM model, with comprehensive infrastructure developed from scratch for code management, development, testing, and analysis, and 4x performance improvement through specific tuning efforts targeting Knight's Landing (KNL) architecture. Following efforts upgrade E3SM from version 1 to version 2 \cite{golaz2022doe-e3sm2}, producing a model that is nearly twice as fast and with a simulated climate that is improved in many metrics. However, most results are only measured on the AMD CPU platform. With the previous tuned KNL architecture becoming obsolete, accelerations of the entire models are still not yet available. Instead, we see a fully-optimized version of only E3SM's nonhydrostatic dycore, HOMMEXX-NH, with an efficient scaling to 276,000 GPUs, and achieving a performance of 0.97 SYPD (simulated years per day) \cite{bertagna2020performance}. The porting process of HOMMEXX-NH involves the rewriting of the dycore solver from Fortran to the C++ Kokkos \cite{edwards2014kokkos} format. While the Kokkos rewriting leads to portable performance across a number of accelerator architectures, it brings extra learning overhead for the climate Fortran community. 

The Consortium for Small-Scale Modeling (COSMO) model is probably the very first regional model that get fully migrated to a GPU platform \cite{fuhrer2018near}. By using up to 4,888 GPUs of the Piz Daint supercomputer, COSMO achieves a performance of 0.043 SYPD for a near-global 1km resolution configuration. The porting of the physics parts rely on compiler directives. In contrast, the porting of the dynamic solver part is similar to the HOMMEXX-NH work, involving the process of rewriting from Fortran to C++, and the introduction of a C++ based domain specific language called Stencil Loop Language (STELLA) \cite{gysi2015stella}. 

While NICAM was the first to try kilometer-level global simulations, its transition from Earth Simulator to K Computer has not been an easy journey \cite{satoh2017outcomes}. With computations distributed over various kernels, with each kernel being improved through optimization, the performance gain of the total simulation is marginal. In contrast, for the transition of NICAM from K Computer to Fugaku \cite{sato2020co}, a careful codesign process is taken to assure over 100 times performance improvement. On top of the optimized version on Fugaku, a data assimilation (DA) system called NICAM-LETKF (local ensemble transform Kalman filter) manages to scale to 131,072 nodes (6,291,456 cores) with a sustained performance of 79 Pflops for the DA part (with an experiment region configured with different spatial resolutions from 65 km to 3.5 km).

While the above efforts only handle the atmosphere component, a most similar effort to our work of porting and optimizing an entire coupled model is the ICON-Sapphire configuration \cite{hohenegger2023icon-sapphire} of the ICON earth system model \cite{jungclaus2022icon}. With the nonhydrostatic dynamic solver developed in 2015 \cite{zangl2015icon-dycore}, and the porting of the ICON model to GPUs \cite{giorgetta2022icon-gpu} (mainly through OpenACC directives), the research team from MPI-M (Max Planck Institute for Meteorology) has moved consistently towards the goal of an earth system model that can depict global climate events at the scale of a few kilometers. With ICON-Sapphire scaling to 600 AMD nodes (76,800 cores), it achieves a simulation speed of 126 SDPD with a 5-km resolution. 
While ICON-Sapphire demonstrates a fairly fast simulation speed in terms of SDPD on both CPU and GPU platforms, its current implementation omits important parameterization schemes such as deep convection, which is shown to have profound effects even at finer resolutions and with nonhydrostatic formulations \cite{wedi2020baseline,Freitas2020GRL}. 

In contrast to the above projects, our work takes a more non-intrusive approach to migrate one of the most widely used earth system model CESM 2.2 to a Sunway supercomputer equipped with more than 100,000 SW26010P processors, each of which consists of 6 core groups (CGs), and 390 cores in total. Instead of adopting a nonhydrostatic solver and removing certain physics schemes, we choose a progressive approach and retain all the current solvers [with the hydrostatic dycore of CAM-SE (Community Atmospheric Model with the Spectral Element method) ] and full physics schemes in the current CESM configuration, so as to maintain the consistency of scientific results across the resolutions (details in section \ref{sec:grid_hierarchy}, and results demonstrated in section \ref{tab:comparison}). While we do not enforce any algorithmic changes to the current CESM 2.2 codebase, we adopt a grid system with a hierarchy of resolutions, an OpenMP-based offloading framework, a redesigned MPI communication scheme for initialization, as well as a complete chain of tools to facilitate the complete flow of migrating the code. 

Table \ref{tab:comparison} demonstrates our results as compared to the state-of-art efforts discussed above. The ported coupled model manages to scale to 101,800 nodes (\ncore cores), and achieves a simulation speed of \finc SDPD. As far as we know, this is a first version of CESM 2.2, with over 80\% of the entire computation ported and parallelized.  

\subsection{Our Previous Eight-Year Journey}

Our journey of enabling high-resolution and ultra-high-resolution climate modeling on Sunway supercomputers started in July, 2015, when we started porting CESM as a first complex scientific application on Sunway TaihuLight \cite{fu2016sunway}.

After achieving a preliminary porting of CESM onto Sunway TaihuLight, we focus on optimization of CAM. The first-round effort refactors the CAM code (both the dynamic part and the physics part) through loop transformations and OpenACC directives, with only marginal speedup for kernels \cite{fu2016refactoring}. For the second round of efforts, we dived one level down, and took a rewriting of the Fortran dycore part into C (similar to the rewriting of HOMMEXX in \cite{bertagna2020performance}), so as to expose the Sunway Athread interfaces for fine-grained control of parallelization and memory operations. We then achieve significantly improved performance, and provide a simulation speed of around 20 SYPD for a ne30 configuration \cite{fu2017redesigning}. 

Afterwards, we collaborated with NCAR to apply similar ideas to the other components of the model, and to adopt system-level tuning for CESM. The work led to a very first heterogeneous version of CESM, called CESM-HR (25-km atmosphere and 10-km ocean) \cite{zhang2020optimizing}, and generated an unprecedented set of high-resolution earth system simulations with a span of 750 years \cite{chang2020unprecedented}.

While the efforts till then worked out smoothly, further development of CESM-HR faced a dilemma after the CESM 2.2 update. With a significant portion of the code modified through the transition, the integration of existing optimization from CESM 1.3 to 2.2 becomes a task as challenging as a from-scratch project. As a result, we choose to stall the performance tuning, but to continue the development of a ultra-high-resolution version, with only a modest speed of simulating several days per day.

From around three weeks ago, with resources allocated for half-scale and full-scale runs on the 40-million-core Sunway supercomputer, we form a team of developers with rich experience of working with CESM along the eight-year journey. With very limited budget of time, and important lessons learnt from the previous journey, we decide to take a non-intrusive approach that minimizes the potential modifications that we need to make to the code, and with some fortune, manage to report results of of some concrete progress.

\section{Innovations}

\subsection{A Hierarchical Grid System}\label{sec:grid_hierarchy}

In order to construct an ultra-high resolution coupled climate model, we build upon established coupled framework of CESM and incorporate kilometer-resolution grids in both the atmosphere and the ocean components. A series of resolutions is used to construct a hierarchy of atmosphere, ocean, and coupled models.
Specifically, the resolution of nominal 0.25$^\circ$ to 0.06$^\circ$ is constructed for CAM-SE, the atmospheric component of CESM, forming the hierarchy of quasi-uniform grids of NE120, NE240 and NE480.
In particular, the resolution of NE480 reaches 5km, which is quadrupled from our previous work \cite{zhang2020optimizing}.
For the ocean and sea ice component, the grids are constructed with a new tripolar grid generation method based on conformal mapping (TS grid, as in \cite{Xu2021GMD}).
The grid hierarchy is self-embedding, and spans a wide range of ocean mesoscale-resolving resolutions from 0.15$^\circ$ (TS015) to 0.03$^\circ$ (TS003).
At 0.03-deg, the nominal resolution for oceanic region is at 2.4km globally, and 1.4km in polar regions (beyond 60$^\circ{N}$ and 60$^\circ{S}$).
The coupling of the atmosphere and the ocean grids follows the standard protocol for CESM, differentiating between the flux coupling (conservative mapping), the atmospheric momentum forcing (high-order mapping) and that for state variables (bilinear mappings). 

For CAM-SE, the nominal resolution of 5 km is reaching the spatial scale of evident nonhydrostatic effects of atmospheric processes.
However, the model’s effective resolution is usually over 5 to 7 times that of the grid’s native resolution, and it is well beyond the spatial scale relevant to nonhydrostatic processes. Hence, we use the default dynamic core of CAM-SE for all the configurations of the resolution hierarchy in this study. For configurations of even higher resolutions (i.e., approaching 1km), the nonhydrostatic effects should be accounted for, e.g., by using new versions of the dynamic core. 

The resolution hierarchy in both the components of the CESM enables us a progressive development of the final, ultra-high resolution coupled model.
The atmosphere-only and ocean-sea ice coupled runs are configured and tested independently using forced experiments. 
Also, the spun-up status of these components are further used to initialize the fully coupled runs.
More importantly, the resolution hierarchy facilitates cross-resolution studies for model tuning, validation, as well as the resolution/scale-dependent processes \cite{zhang2023toward,Xu2021GMD}.

\begin{figure*}[tb]
\centering
\subfigure[The compiling workflow when using the O2ATH plugin]
{
\includegraphics[width=0.9\textwidth]{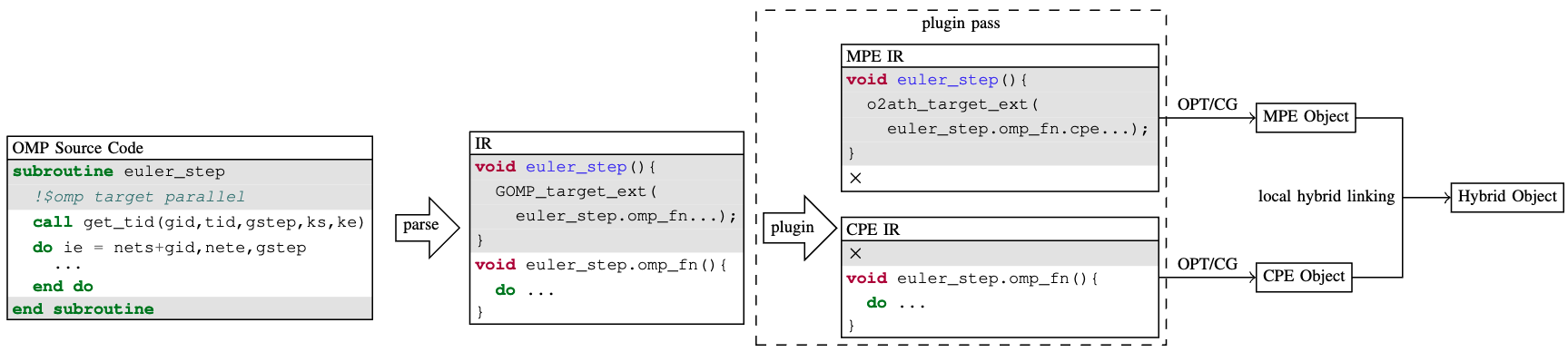}
\label{fig:o2ath-comp}
}
\subfigure[The execution workflow when using the O2ATH runtime library]
{
\includegraphics[width=0.9\textwidth]{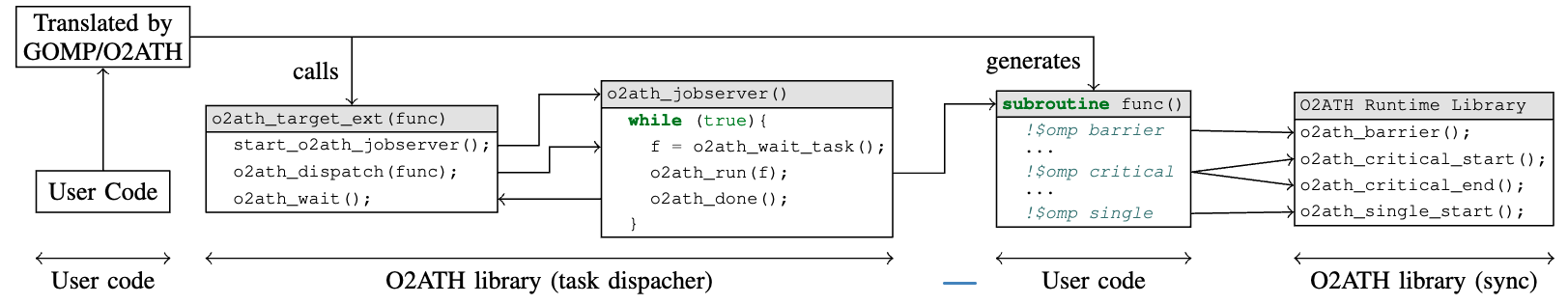}
\label{fig:o2ath-run}
}
\centering
\caption{A detailed example for the compiling and execution workflow of a kernel, when using the O2ATH toolkit.}
\label{fig:hybrid}
\end{figure*}

\subsection{A Non-intrusive Proxy Toolkit O2ATH for OpenMP Offloading to Heterogeneous Manycore Architectures}

While our previous efforts of redesigning CAM involve manual rewriting and tuning for hundreds of thousands of lines of code \cite{fu2017redesigning}, we take a non-intrusive porting strategy in this work, to achieve a better balance between computational performance and consistency of the scientific software code. 

As the Sunway Compiler for the upgraded SW26010P processor is built based on GCC, we can use the compiler to facilitate the generation of interfacing code with GNU OpenMP runtime library, libgomp. Based on the interfacing mechanism, we develop a proxy toolkit, O2ATH, which forwards GOMP library calls to Sunway's Athread Library. 

O2ATH includes a compiler plugin and a runtime library. The plugin determines whether each function should be compiled for CPE or MPE, and eliminate unneeded functions in compiler IR, as shown in Fig.~\ref{fig:o2ath-comp}. The runtime library supports task dispatching, and basic OpenMP barrier/single/critical constructs, as shown in Fig.~\ref{fig:o2ath-run}.

The most challenging obstacle that stop us from implementing OpenMP features through CPE threads is the synchronization construct, such as barrier, single, and critical constructs. These constructs are usually used to grant data consistency. As the cache coherency is not fully maintained by CPEs, we add explicit cache flush operations to support these constructs.

Another problem is the different memory model of the SW26010P heterogeneous manycore processor, as compared to traditional OpenMP offloading targets. The size of LDM is smaller than GDDR or HBM by orders of magnitude. Our current solution is to use LDM as CPE stack when the size of LDM is large enough. Otherwise, we place the stack in the CPE's private memory.

While the the hybrid-LTO feature is not usable in swgcc, we write a wrapper script to compile one file twice to generate CPE machine code.
The usage of O2ATH is like SWACC, but the compiling speed is about 10x faster, that is, we can recompile full CESM code in 10 minutes instead of 2 hours. Furthermore, the O2ATH-based optimization does not prevent us from low-level optimizations.

\begin{table*}
\centering
\caption{Parallelization strategies for different component models in CESM.}\label{tab:opts}
\def\x{$\times$}
\resizebox{1\linewidth}{!}{
\begin{tabular}{cccccl}\hline
    \multicolumn{2}{c}{Component}  & Loop Form & Typical Dimension per CG & \multicolumn{1}{c}{Strategy } \\\hline
    \multirow{2}{*}{CAM}  & Dynamics & nelem\x pver\x np\x np & $4\times32\times4\times4$ & Parallel on nelem\x pver \\\cline{2-5}
         & Physics  & nchunks \x ncols\x pver & $36\times 1 \times 32$ & Parallel on nchunks\\\hline
    \multirow{2}{*}{POP}  & VMIX \& clinic \&etc     & mxblk\x nlayer\x nyblk \x nxblk  & $1\times 60\times 56 \times 10$& Parallel on nyblk\\\cline{2-5}
         & HMIX \& etc    & mxblk\x nlayer\x nyblk \x nxblk  & $1\times 60\times 56 \times 10$& Parallel on nlayer\\\hline
    CICE & EVP \& dEdd   & mxblk\x ncat\x nlayer \x nyblk \x nxblk  & $32\times 5\times 8 \times 4\times4$ & Parallel on mxblk\\\hline
\end{tabular}
}
\end{table*}

\subsection{A Wide Adoptation of O2ATH to Major Component Models }

With the OpenMP directive enabled offloading to the CPEs, we can then apply appropriate strategies to achieve suitable parallelization in different component models. The scheme works fine across the atmosphere component CAM, the ocean component POP, and the sea-ice component CICE. The land component CLM is not considered at this time, due to our limited time budget and its relatively small compute cost in the entire coupled model.

The wide coverage of the approach proves to be an very important metric for the case of porting an earth system model, which is probably one of the few parallel programs that are clearly doomed by Amdahl's Law. With numerous kernels that cover different parts of climate scientists' understanding of various change factors, optimizing only a part of the model generally provide marginal performance benefits \cite{satoh2017outcomes}. Applying O2ATH, we can design homogeneous strategies for each major computation pattern in CESM, so as to achieve a well-covered performance boost. 

A brief introduction of these strategies is shown in TABLE \ref{tab:opts}. For the 64 CPEs within each CG, we generally pick a variable that both gives enough parallelism and provides a good balance across compute, memory access, and loop layout. For example, for the dycore part of CAM, we parallelize over elements and vertical layers; while for the physics part of CAM, we parallelize over chunks and set columns per chunk to 1. For POP, we apply different strategies for vertical and horizontal mixers. For CICE, we  parallelize over the blocks.

\subsection{Boosting the Initialization Stage}

The long initialization time of CESM has been a historical issue for the team. While a six-hour initialization sounds reasonable for a several-month climate modeling experiment, such a cycle soon becomes the bottleneck during a porting process, especially for a porting to the full scale of a 40-million-core supercomputer. On the other hand, many components cannot scale to hundreds of thousands of processes in the original version, and thus optimization of initialization stage is imperative for enabling full-machine run.

While our previous effort \cite{zhang2020optimizing,li2021enabling} has already alleviated the issue for large-scale runs on Sunway TaihuLight, the increase of the parallel scale from around 80,000 MPI processes to 610,800 MPI processes, clearly brings new problems.

To reduce the initialization time, and to maximize the number of experiments within a limited amount of time, we have identified the following effective techniques:
\begin{itemize}
\item We reform the MPI communication pattern at the initialization stage. The rearrange communication in PIO and MCT is replaced by a hierarchical AlltoAllw, and the blockwise communication in CICE is replaced by a packed multi-stage Gatherv communication. Both optimizations effectively reduce the memory requirement and RDMA queue resources when scaling to the full machine. 
\item We reduce time complexity of the original mapping algorithm from physical node ID to MPI process Rank from $O(N^2)$ to $O(Nlog(N))$ through the quick sort algorithm of fixed-length strings.
\item  We balance the I/O overhead and the communication overhead of PIO under large-scale concurrency by adjusting the file system layout and the number of I/O processes used (Based on the varying levels of parallelism and initialization data volumes within different models in CESM, the Lustre parameters (stripe\_count, stripe\_size) are adjusted from their default values of (1, 1 MB) to configurations of ([4, 8, 16, 32], [1MB, 4MB]). And, every 1000 processes share one I/O process). 
\item For LND part, we reduce time complexity of the smp-level clumps distribution from $O(N^2)$ to $O(N)$. We also store the initialization files of CLM.
\end{itemize}

\begin{table}
\centering
\caption{Results of Initialization Optimization}
\label{tab:init}
\begin{tabular}{cccccc}
\toprule
\multirow{2}{*}{Version} & \multirow{2}{*}{Init time} & \multicolumn{4}{c}{Scale}\\\cline{3-6}
 &  & ATM & ICE & LND & OCN  \\
\midrule
Original & 6 hours & 200k & 50k & 50k & 100k \\
Optimized & 35 minus & 460k & 460k & 460k & 200k \\
\bottomrule
\end{tabular}
\end{table}

As TABLE \ref{tab:init} shown, with all strategies listed above we manage to reduce the initialization time from 6 hours to 35 minutes for full-machine run of kilometer-level coupled modeling. For 25 kilometers resolution with weak scaling, the initialization time is even less than 10 minutes. Meanwhile, the maximum runnable scale of each component is significantly increased, which make full-machine run realizable.

\subsection{Other Tools}

At the very beginning of the project, we have made the strategy to minimize the manual code modifications that we need to make, and to maximize the benefits that can be brought by tools. Therefore, a large amount of our efforts are spent on improving existing tools and developing new tools to facilitate this three-week porting exploration. 

We list the major tools and their main features as follows:
\begin{itemize}
\item SWMU: memory-related debugging and wrap-based memory usage analysis;
\item SWCallgraph: generating and analyzing static call graph via binary scanning, an extremely useful tool for debugging;
\item SWLU: providing performance profiling results (updated to cover most aspects that we need to tend to, such as compute time for MPE and CPE, communication time, I/O time, etc.) based on sampling;
\item BitwiseCheck: providing correctness verification between MPE version and ported MPE+CPE version.
\item CPE-divert: a compiler plugin that handles the conversion from MPE virtual function pointers to CPE virtual function pointers. Combining CPE-divert and O2ATH, we can directly call the class methods in physics schemes when offloading to CPEs.
\item libvnest: library for providing support in nested vertical layer parallelism in dynamics core, including loop index handling and solving data dependency by transposing or RMA based methods.
\end{itemize}

\section{How Performance Was Measured}

\subsection{Major Performance Metrics}

In terms of timing, the performance is measured using the average time recorded for running the same case for three times. For the climate modeling scenario, the speed that we can achieve to performance simulation experiment is obviously a more important metrics to consider. For most component model and coupled model results, we describe the speed of simulation using either SYPD (simulated years per day) or SDPD (simulated days per day, when we are too far away from a year) metrics. Note that, these metrics need to be evaluated with the resolution configurations.

\subsection{Hardware Platform Details}

Our experiments are performed on the Sunway supercomputer that precedes Sunway TaihuLight. A major update is the SW26010P processor used in the supercomputer, an upgraded version of the previous SW26010 processor.

Each SW26010P processor consists of 6 core groups (CGs), connected via a network-on-chip (NoC). Each CG consists of one management processing element (MPE) and 64 computing processing elements (CPEs) organized as an 8$\times$8 array, with 390 cores in total. MPE and CPE provide support for 256-bit and 512-bit SIMD instruction, respectively. Each computing node has a theoretical peak performance of 14 Tflops.

From a memory hierarchy point of view, SW26010P has a total of 96GB DDR4 memory, distributed across 6 CGs, with 16GB for each CG with a bandwidth of 51.2 GB/s. Each CPE has a 32KB L1 instruction cache and a 256KB user-controlled scratchpad memory, also referred to as local data memory (LDM). Different from its predecessor, LDM in SW26010P can be partly configured as a hardware-controlled local data cache to optimize memory access.

Data transfer between global memory and LDM can be accomplished by direct memory access (DMA), which is the recommended way for batch data transfer. Communications among CPEs in the same CG are achieved by remote memory access (RMA).  
Four adjacent CPEs are connected to the same router where data exchange is realized in CG.
The theoretical bandwidth of DMA and RMA is 307 GB/s and 460 GB/s, for transferring chunks of contiguous data.

\begin{figure}[t]                                    
    \centering 
    \includegraphics[width=0.48\textwidth]{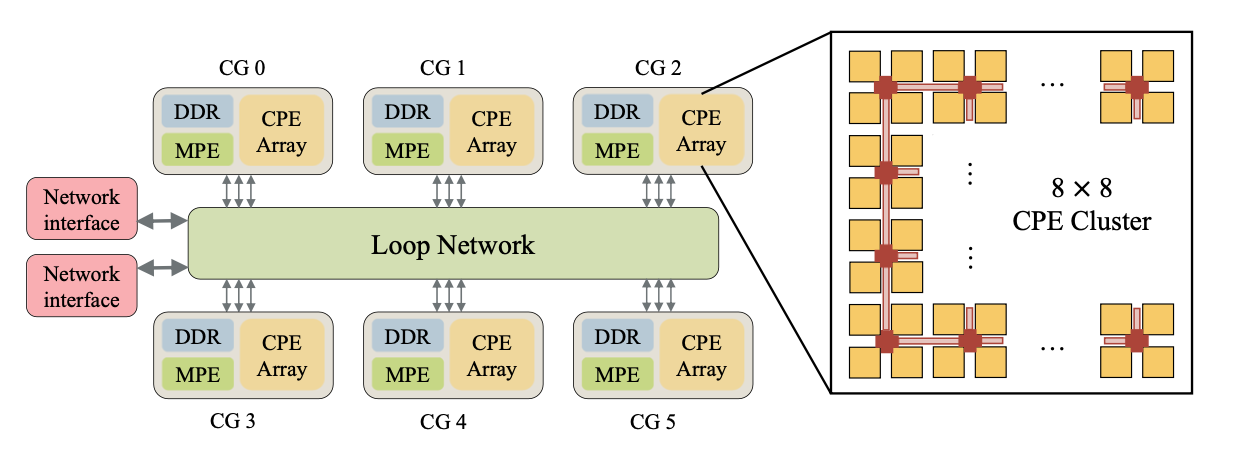}
    \caption{SW26010P processor.}
    \label{fig:sw_pro}   
\end{figure} 

\section{Performance Results}

\begin{figure}[t]                                    
    \centering 
    \includegraphics[width=0.48\textwidth]{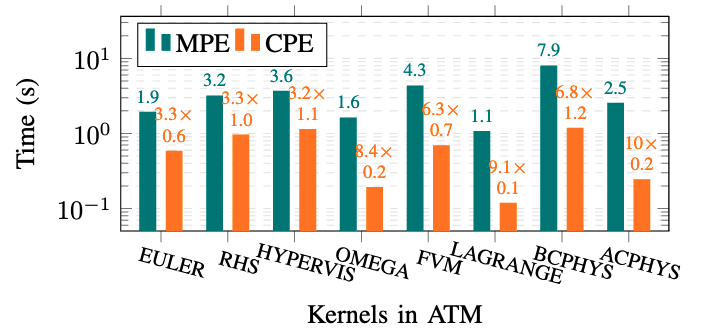}
    \caption{Performance improvements on CPEs for major kernels in the atmosphere component CAM.}
    \label{fig:part-kernel-ATM}   
\end{figure} 

\begin{figure}[t]                                    
    \centering 
    \includegraphics[width=0.48\textwidth]{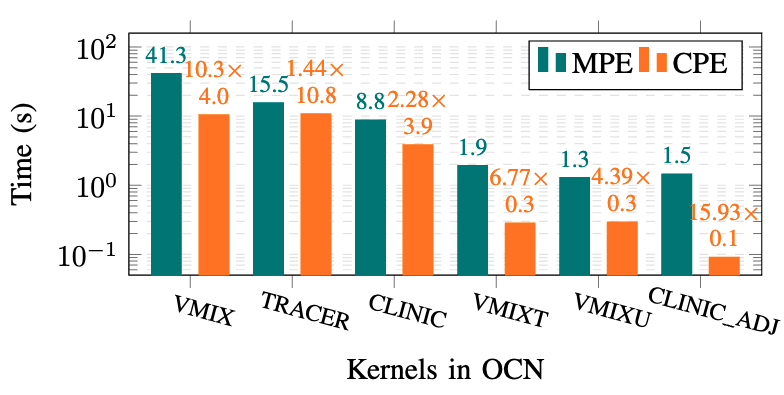}
    \caption{Performance improvements on CPEs for major kernels in the ocean component POP.}
    \label{fig:part-kernel-OCN}   
\end{figure}

\begin{figure}[tb]                                    
    \centering 
    \includegraphics[width=0.48\textwidth]{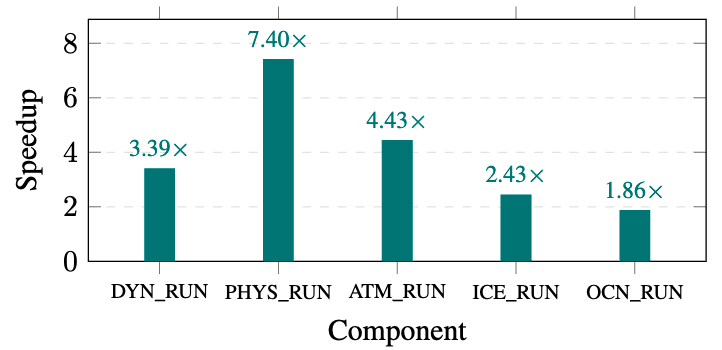}
    \caption{Performance improvements on CPEs for the component models.}
    \label{fig:whole-kernel}   
\end{figure} 

\subsection{Accelerations of Major Kernels on CPEs}
The first step to take advantage of the heterogeneous many-core processor is to parallelize the computation over the array of CPEs. With the wide adoption of O2ATH toolkit, most kernels are parallelized from MPE to CPEs without extra coding efforts. The overview of performance improvements of important kernels in CAM and POP are shown in Fig.~\ref{fig:part-kernel-ATM} and Fig.~\ref{fig:part-kernel-OCN}. The improvement is mostly factorized in two aspects: communication and parallelism.

Most kernels in the dycore of CAM contain boundary exchange among elements. Kernels like EULER, RHS, HYPERVIS suffer from Amdahl's Law, and do not show good speedups. Benefited from CPE parallelization, kernels like OMEGA, FVM and LAGRANGE can gain speedups ranging from $6\times$ to $9\times$, and almost reach the theoretical memory bandwidth upper bound. Another effective strategy in the dycore is that we use prefix sum tool from libvnest to resolve the vertical layer dependency of kernel OMEGA, thus removing the performance impact from data dependency.

One highlight worth mentioning is that our workflow facilitates the porting of the entire physics scheme on to heterogeneous CPE arrays, leaving most code unchanged. Before-coupling and after-coupling physics (BCPHYS, ACPHYS) gain speedups up to $6.8\times$ and $10\times$. A possible reason for the relatively lower speedup of BCPHYS is probably the more complex code and the resulting poor cache behaviors and bad branch predictions. 

For kernels in POP2, by partitioning horizontal blocks into a "thin and tall" shape, both vertical parallelism and horizontal parallelism inside blocks can gain promising speedups.

Fig.~\ref{fig:whole-kernel} shows the performance improvements of entire component models, with the dynamic part and physics part of CAM also demonstrated. The physics part of CAM shows the highest speedup, as no communication is involved. In contrast, CICE is largely constrained by the communication bottleneck. Moreover, with the reduction of the atmosphere timestep size, the computation cost of CICE is also increasing since it is tightly coupled with CAM, making CICE even slower than CAM in ultra-high-resolution configuration. The speedup of POP and CICE on CPEs are 1.86 and 2.43 times respectively.

\subsection{Weak and Strong Scaling Results of CAM and POP}

\def\cplsdpd{20}
\begin{figure}[t]                                    
    \centering 
    \includegraphics[width=0.48\textwidth]{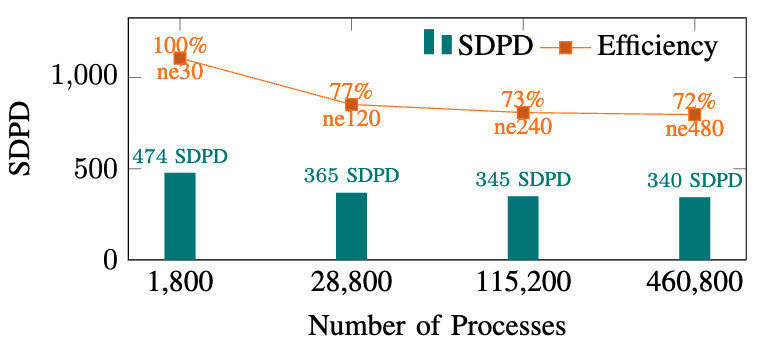}
    \caption{Weak scaling results of CAM. ne30, ne120, ne240, and ne480 correspond to 100-km, 25-km, 10-km, and 5-km resolution respectively.}
    \label{fig:weak_atm}   
\end{figure} 

\begin{figure}[t]                                    
    \centering 
    \includegraphics[width=0.48\textwidth]{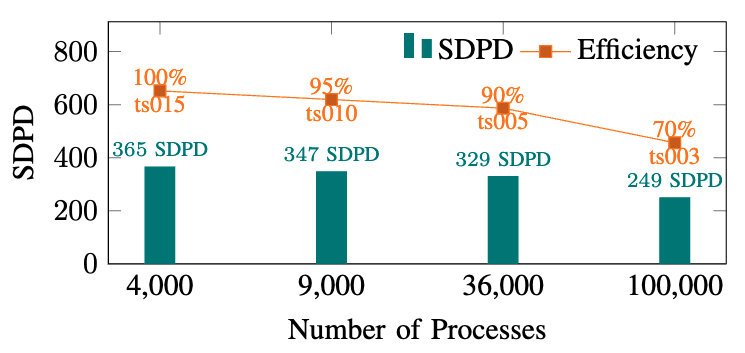}
    \caption{Weak scaling results of POP. ts015, ts010, ts005, and ts003 refer to 15-km, 10-km, 5-km and 3-km resolutions respectively.}
    \label{fig:weak_ocn}   
\end{figure}

The weak scaling results of CAM and POP are shown in Fig.~\ref{fig:weak_atm} and Fig.~\ref{fig:weak_ocn}. The result is acquired by setting the same timestep between low-resolution and high-resolution cases, and keep the same ratio of number of grid points to processor. The parallel efficiency of CAM drops to 77\%, with the communication changing from intra-cabinet (ne30) to inter-cabinet (ne120). The parallel efficiency further drops to around 72\% when we gradually expand to more than half of the full machine. The parallel efficiency of POP shows similar drops for the switching of communication patterns, and increase of parallel scale, but is generally better than CAM as it consumes a smaller scale of parallel nodes.

\begin{figure}[t]                                    
    \centering 
    \includegraphics[width=0.48\textwidth]{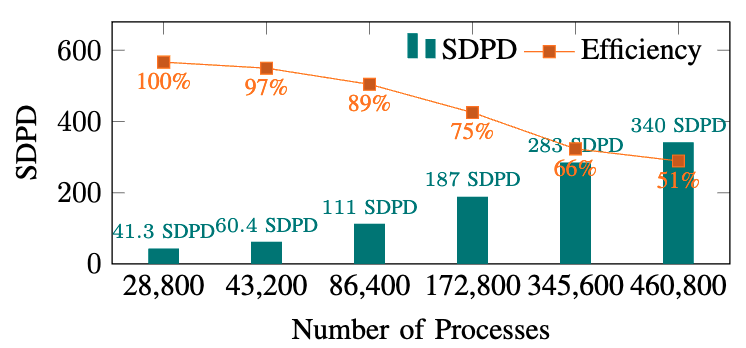}
    \caption{Strong scaling results of CAM, for the 5-km resolution.}
    \label{fig:strong_atm}   
\end{figure} 

\begin{figure}[t]                                    
    \centering 
    \includegraphics[width=0.48\textwidth]{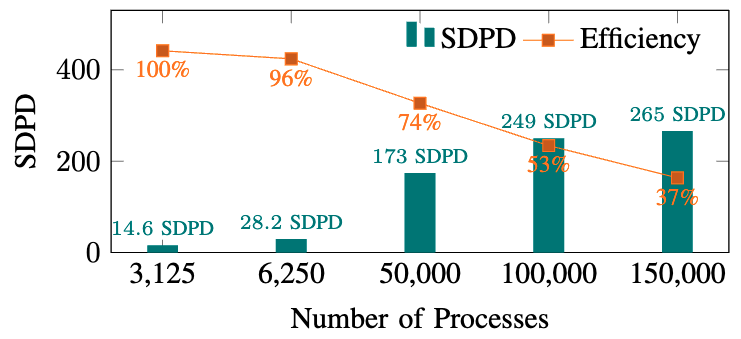}
    \caption{Strong scaling results of POP, for the 3-km resolution.}
    \label{fig:strong_ocn}   
\end{figure}

Strong scaling results are shown in Fig.~\ref{fig:strong_atm} and \ref{fig:strong_ocn}. The starting point of 5-km CAM uses 28,800 processes (slightly more than 17,525 minimum requirement), and achieves a speed of 41.28 SDPD. The scaling efficiency of CAM for 460,800 processes is about 47\%. 3-km POP scales from 14.6 SDPD in 3,125 process to \fino SDPD in 150,000 processes, and achieves 37\% strong scaling efficiency.

\begin{figure*}[tb]
    \centering
    \includegraphics[width=0.95\textwidth]{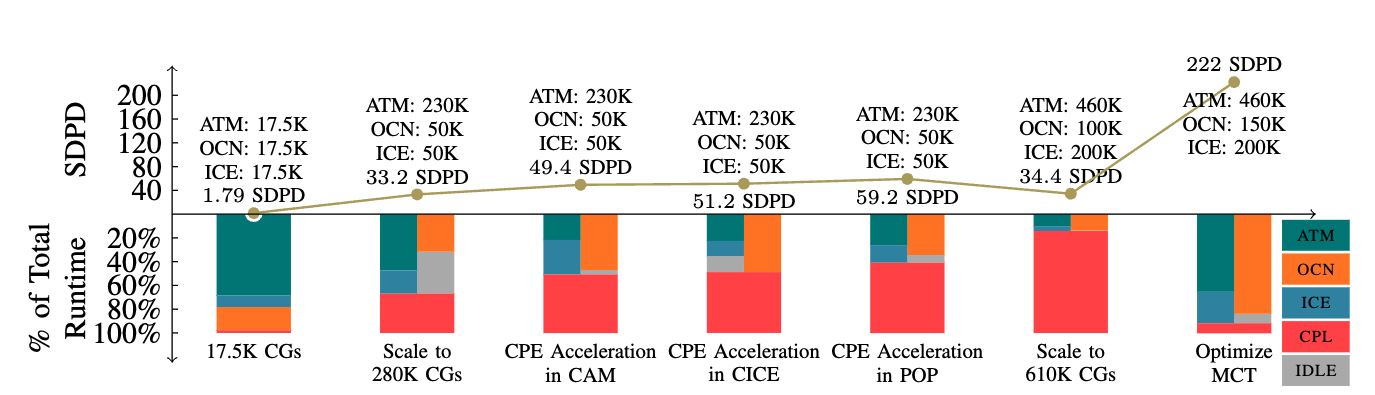}
    \caption{Important performance improvement milestones for large-scale run of the coupled model. ATM, OCN, ICE, CPL, and IDLE blocks refer to the percentage in total run time, taken by the atmosphere component CAM, the ocean component POP, the sea-ice component CICE, the coupler, and the idling of processes waiting for others (indicating load-balancing issues). The curve corresponds to the simulation speed of each configuration, described in SDPDs (simulated days per day). Number of CGs used in each component is described near the curve.}\label{fig:milestone-model}
\end{figure*}
\begin{figure}[t]                                    
    \centering 
    \includegraphics[width=0.48\textwidth]{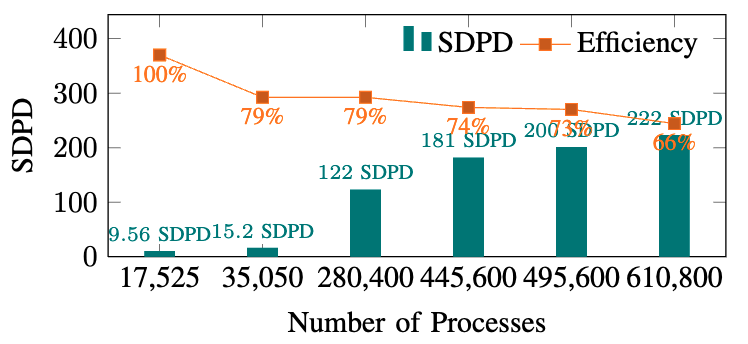}
    \caption{Strong scaling results of the coupled model.}
    \label{fig:strong_coupled}   
\end{figure} 

\subsection{Large-Scale Runs of the Coupled Model}

Benefited from the slow-varying characteristic of ocean, the timestep of POP is much larger than CAM, and it can reach equivalent performance with much fewer computation nodes. We firstly try our best to scale CAM to the full system, and record the performance to determine a matching POP scale. The final performance of CAM on the new Sunway supercomputer is \fina SDPD (close to 1 SYPD). After that, we find that POP can reach similar simulation speed with around 150,000 CGs. We then decide the final parallelization partition between CAM and POP for the coupled model.

Fig.~\ref{fig:milestone-model} illustrates important performance improvement milestones, for tuning of large-scale runs. We start from a minimum set of 17,525 processes to support 5-km atmosphere and 3-km ocean coupled modeling. Scaling to 280,400 processes (around half of the machine), we are achieving a speed of 33.2 SDPD, even without using CPEs. By switching on the CPE acceleration in CAM, CICE, and POP, we are gradually increasing towards 59.2 SDPD. After solving the connection number limit problem by using a layered collective MPI communication library, we can scale the model to 610,800 CGs, but the speed drops to 34.4 SDPD as the CPL performance becomes the bottleneck. SWLU identifies MPI all-to-all communication as the major reason for CPL's poor performance, and we reconfigured MCT to use send-recv scheme in rearrangement. The final performance of our implementation is \finc SDPD. Another issue worth mentioning is our underestimation of CICE's computation cost in high-resolution cases. In the extreme case, CICE needs more resources than POP to achieve satisfactory performance. 

Fig.\ref{fig:strong_coupled} shows the strong scaling result of the coupled 5-km atmosphere 3-km ocean model. Starting from a minimum case to support such a resolution, the full-machine case still maintains a parallel efficiency of around 66\%.

\subsection{Scientific Results}

%\textcolor{red}{xusm: text and figures, .5p. It would be good if we can demonstrate that we are doing science better than ICON since we are probably paying more computational costs}

The ultra-high resolution configuration of the model is initialized with spun-up status of the 0.25$^\circ$ coupled experiment and that based on the atmospherically forced long-term run, respectively. 
The baroclinic time steps of the atmosphere and the ocean components are at 30$s$ and 90$s$, respectively, with the coupling frequency of 180$s$. 
This configuration, together with others of the resolution hierarchy, are run for about 1 year and the results after initial spin-up processes are used for cross-resolution inter-comparisons and the study various processes.
Key characteristics of the coupled climate system are all well characterized with hierarchy of model resolutions, including tropical cyclones (TC), mesoscale/submesoscale modulated air-sea interactions, topography/bathymetry effects on geophysical fluid. 
For example, slightly stronger TCs are simulated with 5km-resolution resolution than 25km resolution, but the modeled structure of the TCs, including the eye-wall and wind/precipitation bands is much finer in the former \cite{zhang2023toward}.
Also, storm track is much stronger, due to finer sea-surface temperature gradients and ensuing air-sea interactions enabled at submesoscale-permitting resolution of TS003. 

\begin{figure}[!htbp]
\centering
\includegraphics[width=0.5\linewidth]{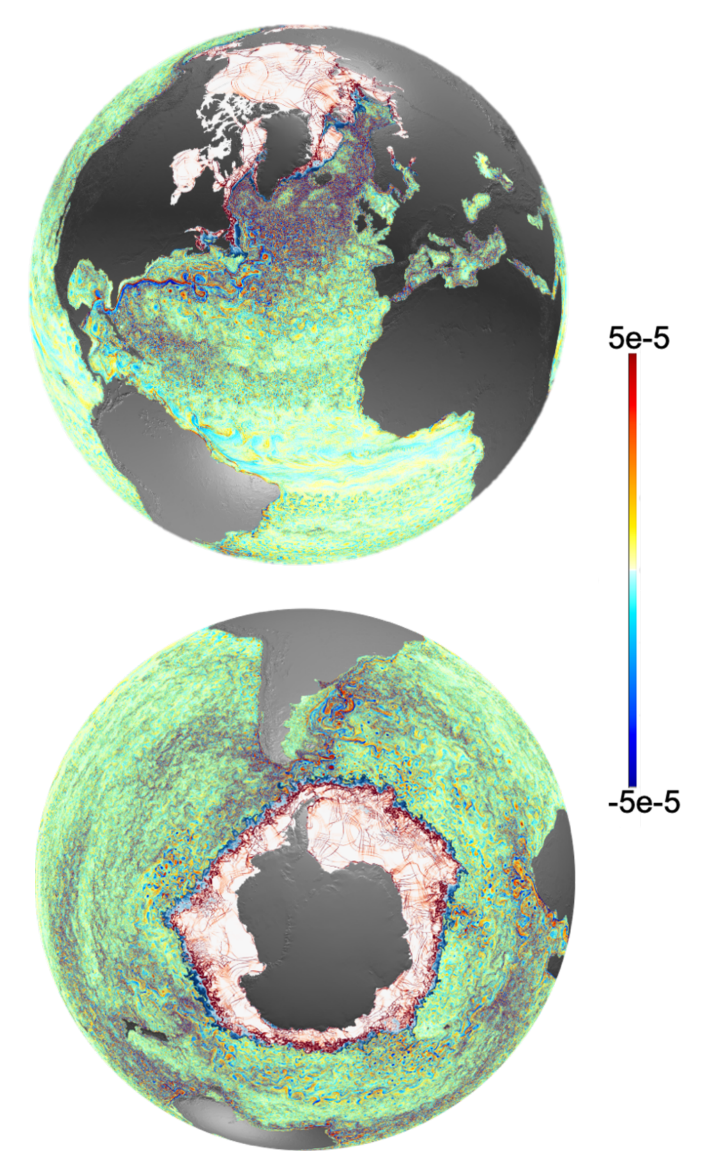}
\caption{Ocean surface relative vorticity ($\zeta$) and sea ice concentration/deformation of wintertime northern (top) and southern (bottom) hemisphere for the ultra-high resolution coupled run (NE480 and TS003).}\label{fig:so_zeta}
\end{figure}

Fig. \ref{fig:so_zeta}. shows the snapshot of ocean surface’s vorticity and sea ice deformation during winter for both hemispheres for the coupled run with NE480 (atm) and TS003 (ocn). 
Highly ageostrophic, submesoscale-rich oceanic flow is prevalent in key air-sea interaction regions of subtropical gyres and ACC (Antarctic circumpolar current).
Linear kinematic features, which are unique to sea ice dynamics, arise due to multi-scale, fractal plastic deformations, which have profound effect on moisture and buoyancy fluxes across the air-sea interface.
As a unique characteristics of the our approach, the resolution hierarchy in both the atmospheric and the oceanic components enables us the cross-resolution/scale study of the modeled processes.
With the resolution increase from 0.15$^\circ$ to 0.03$^\circ$, the atmospherically forced ocean simulations witness over 90\% increase in the the total ocean kinematic energy. 
Although at 0.05$^\circ$ (4km global average) the model is capable to resolve certain submesoscale processes, finer structures such as ocean fronts are much better characterized at 0.03$^\circ$.
Coupling with the atmospheric model with NE480 introduces much higher spatial and temporal variability for the ocean's forcing, resulting in further increase of the overall kinetic energy by another 12\% and that of the surface ocean by 21\%. 

Continued numerical integration and further analysis of the model output is currently underway. 
In particular, we plan to optimize the model's performance by fully utilizing the multiple resolutions/scales of the resolution hierarchy. 
The air-sea coupling processes, as well as the reduction of the long-standing biases of climate models, are among the topics of study with longer, multi-year's simulation results.

%%%%%%%%%%%%%%%%%%%%%%%%%%%%%%%%%%%%%%%%%%%%%%%%%%%%%%%%%%%%%
\section{Implication}

%In our previous submissions, we have this tradition of quoting one important sentence from a traditional Chinese literature to provide a neat alignment and suitable echo with the main idea of the article. Usually, the sentence came out early at the stage of writing the introduction part. This time, it waits till when we conclude the paper.

Recorded in the Analects of Confucius \cite{waley2005analects}, Confucius once mentioned to his student Zigong: ``If an engineer wants to improve his work, he needs to improve his tool first." Translated to the context of software development, this sentence still brings valuable insights. Especially for complex software as in the case of a million-line earth system model, we need tools that can not only facilitate the migration of software to a more powerful but architecturally different platform, and but also support the transfer of climate system understanding to a new version of the software that still accommodates most of the related community. In other words, we believe climate modeling is a multidisciplinary subject that requires both scientific expertise and engineering improvements. Along the way towards the ultimate understanding and accurate prediction of the system, we need software developments that do not only target performance of the machine but also the productivity of the people who interact with the machine.

Following such a philosophy, throughout this three-week project, we adopt methods and tools that can support a non-intrusive and widely-effective porting of CESM 2.2. We demonstrate that we manage to improve the simulation speed from {\stc SDPD to \finc SDPD}, making it possible for climate modellers to perform kilometer-resolution experiments at multi-year or even multi-decadal scales.

With further improvement of resolutions, kilometer-scale coupled models will serve as the backbone for various scientific research and key applications in climate change. 
They promise much better characterization of the multi-scale atmospheric, oceanic, as well as coupled processes, such as much finer simulation of the energy cycle, air-sea interactions at frontal scales, etc.
The knowledge, as well as the data produced, will constitute an indispensable source of information for climate model development, future projection, and climate mitigation practices. 

\bibliographystyle{unsrtnat}
\bibliography{./refs-cesm}

\begin{thebibliography}{42}
\providecommand{\natexlab}[1]{#1}
\providecommand{\url}[1]{\texttt{#1}}
\expandafter\ifx\csname urlstyle\endcsname\relax
  \providecommand{\doi}[1]{doi: #1}\else
  \providecommand{\doi}{doi: \begingroup \urlstyle{rm}\Url}\fi

\bibitem[Yang et~al.(2002)Yang, Braeuning, Johnson, and Yafeng]{yang2002general-ChinaTempReconstruction}
Bao Yang, Achim Braeuning, Kathleen~R Johnson, and Shi Yafeng.
\newblock General characteristics of temperature variation in china during the last two millennia.
\newblock \emph{Geophysical research letters}, 29\penalty0 (9):\penalty0 38--1, 2002.

\bibitem[Zhang et~al.(2008)Zhang, Cheng, Edwards, Chen, Wang, Yang, Liu, Tan, Wang, Liu, et~al.]{zhang2008test-monsoon}
Pingzhong Zhang, Hai Cheng, R~Lawrence Edwards, Fahu Chen, Yongjin Wang, Xunlin Yang, Jian Liu, Ming Tan, Xianfeng Wang, Jinghua Liu, et~al.
\newblock A test of climate, sun, and culture relationships from an 1810-year chinese cave record.
\newblock \emph{science}, 322\penalty0 (5903):\penalty0 940--942, 2008.

\bibitem[Zhang et~al.(2010)Zhang, Li, Ku, and Lu]{zhang2010linking}
DeEr Zhang, Hong-Chun Li, Teh-Lung Ku, and LongHua Lu.
\newblock On linking climate to chinese dynastic change: Spatial and temporal variations of monsoonal rain.
\newblock \emph{Chinese Science Bulletin}, 55\penalty0 (1):\penalty0 77--83, 2010.

\bibitem[Guanzhong(2018)]{guanzhong2018romance}
Luo Guanzhong.
\newblock \emph{The romance of the three kingdoms}.
\newblock Penguin UK, 2018.

\bibitem[Platzman(1979)]{platzman1979eniac}
George~W Platzman.
\newblock The eniac computations of 1950—gateway to numerical weather prediction.
\newblock \emph{Bulletin of the American Meteorological Society}, 60\penalty0 (4):\penalty0 302--312, 1979.

\bibitem[McGuffie and Henderson-Sellers(2001)]{mcguffie2001forty}
K~McGuffie and A~Henderson-Sellers.
\newblock Forty years of numerical climate modelling.
\newblock \emph{International Journal of Climatology: A Journal of the Royal Meteorological Society}, 21\penalty0 (9):\penalty0 1067--1109, 2001.

\bibitem[Dickinson(1982)]{dickinson1982optimizing}
A~Dickinson.
\newblock Optimizing numerical weather forecasting models for the cray-1 and cyber 205 computers.
\newblock \emph{Computer Physics Communications}, 26\penalty0 (3-4):\penalty0 459--468, 1982.

\bibitem[Mitchell(1990)]{mitchell1990genius}
Russell Mitchell.
\newblock The genius: Meet seymour cray, father of the supercomputer.
\newblock \emph{Business Week}, 3157\penalty0 (30):\penalty0 80--88, 1990.

\bibitem[Chervin(1988)]{chervin1988relationship}
Robert~M Chervin.
\newblock On the relationship between computer technology and climate modelling.
\newblock \emph{Physically-Based Modelling and Simulation of Climate and Climatic Change: Part 2}, pages 1053--1068, 1988.

\bibitem[Chervin and Semtner~Jr(1990)]{chervin1990ocean}
Robert~M Chervin and Albert~J Semtner~Jr.
\newblock An ocean modelling system for supercomputer architectures of the 1990s.
\newblock In \emph{Climate-Ocean Interaction}, pages 87--95. Springer, 1990.

\bibitem[Semtner~Jr and Chervin(1992)]{semtner1992ocean}
Albert~J Semtner~Jr and Robert~M Chervin.
\newblock Ocean general circulation from a global eddy-resolving model.
\newblock \emph{Journal of Geophysical Research: Oceans}, 97\penalty0 (C4):\penalty0 5493--5550, 1992.

\bibitem[Shingu et~al.(2002)Shingu, Takahara, Fuchigami, Yamada, Tsuda, Ohfuchi, Sasaki, Kobayashi, Hagiwara, Habata, et~al.]{shingu200226}
Satoru Shingu, Hiroshi Takahara, Hiromitsu Fuchigami, Masayuki Yamada, Yoshinori Tsuda, Wataru Ohfuchi, Yuji Sasaki, Kazuo Kobayashi, Takashi Hagiwara, Shin-ichi Habata, et~al.
\newblock A 26.58 tflops global atmospheric simulation with the spectral transform method on the earth simulator.
\newblock In \emph{Proceedings of the 2002 ACM/IEEE conference on Supercomputing}, pages 1--19. Citeseer, 2002.

\bibitem[Michalakes et~al.(2007)Michalakes, Hacker, Loft, McCracken, Snavely, Wright, Spelce, Gorda, and Walkup]{michalakes2007wrf}
John Michalakes, Josh Hacker, Richard Loft, Michael~O McCracken, Allan Snavely, Nicholas~J Wright, Tom Spelce, Brent Gorda, and Robert Walkup.
\newblock Wrf nature run.
\newblock In \emph{Proceedings of the 2007 ACM/IEEE conference on Supercomputing}, pages 1--6, 2007.

\bibitem[Fu et~al.(2016{\natexlab{a}})Fu, Liao, Xue, Wang, Chen, Gu, Xu, Ding, Wang, He, et~al.]{fu2016refactoring}
Haohuan Fu, Junfeng Liao, Wei Xue, Lanning Wang, Dexun Chen, Long Gu, Jinxiu Xu, Nan Ding, Xinliang Wang, Conghui He, et~al.
\newblock Refactoring and optimizing the community atmosphere model (cam) on the sunway taihulight supercomputer.
\newblock In \emph{SC'16: Proceedings of the International Conference for High Performance Computing, Networking, Storage and Analysis}, pages 969--980. IEEE, 2016{\natexlab{a}}.

\bibitem[Fu et~al.(2017)Fu, Liao, Ding, Duan, Gan, Liang, Wang, Yang, Zheng, Liu, et~al.]{fu2017redesigning}
Haohuan Fu, Junfeng Liao, Nan Ding, Xiaohui Duan, Lin Gan, Yishuang Liang, Xinliang Wang, Jinzhe Yang, Yan Zheng, Weiguo Liu, et~al.
\newblock Redesigning cam-se for peta-scale climate modeling performance and ultra-high resolution on sunway taihulight.
\newblock In \emph{Proceedings of the international conference for high performance computing, networking, storage and analysis}, pages 1--12, 2017.

\bibitem[Leung et~al.(2020)Leung, Bader, Taylor, and McCoy]{leung2020introduction-e3sm}
L~Ruby Leung, David~C Bader, Mark~A Taylor, and Renata~B McCoy.
\newblock An introduction to the e3sm special collection: Goals, science drivers, development, and analysis.
\newblock \emph{Journal of Advances in Modeling Earth Systems}, 12\penalty0 (11):\penalty0 e2019MS001821, 2020.

\bibitem[Sch{\"a}r et~al.(2020)Sch{\"a}r, Fuhrer, Arteaga, Ban, Charpilloz, Di~Girolamo, Hentgen, Hoefler, Lapillonne, Leutwyler, et~al.]{schar2020kilometer}
Christoph Sch{\"a}r, Oliver Fuhrer, Andrea Arteaga, Nikolina Ban, Christophe Charpilloz, Salvatore Di~Girolamo, Laureline Hentgen, Torsten Hoefler, Xavier Lapillonne, David Leutwyler, et~al.
\newblock Kilometer-scale climate models: Prospects and challenges.
\newblock \emph{Bulletin of the American Meteorological Society}, 101\penalty0 (5):\penalty0 E567--E587, 2020.

\bibitem[Lin et~al.(2020)Lin, Huang, Liang, Qin, Xu, Huang, Xu, Liu, Wang, Peng, et~al.]{lin2020community}
Yanluan Lin, Xiaomeng Huang, Yishuang Liang, Yi~Qin, Shiming Xu, Wenyu Huang, Fanghua Xu, Li~Liu, Yong Wang, Yiran Peng, et~al.
\newblock Community integrated earth system model (ciesm): Description and evaluation.
\newblock \emph{Journal of Advances in Modeling Earth Systems}, 12\penalty0 (8):\penalty0 e2019MS002036, 2020.

\bibitem[Chang et~al.(2020)Chang, Zhang, Danabasoglu, Yeager, Fu, Wang, Castruccio, Chen, Edwards, Fu, et~al.]{chang2020unprecedented}
Ping Chang, Shaoqing Zhang, Gokhan Danabasoglu, Stephen~G Yeager, Haohuan Fu, Hong Wang, Frederic~S Castruccio, Yuhu Chen, James Edwards, Dan Fu, et~al.
\newblock An unprecedented set of high-resolution earth system simulations for understanding multiscale interactions in climate variability and change.
\newblock \emph{Journal of Advances in Modeling Earth Systems}, 12\penalty0 (12):\penalty0 e2020MS002298, 2020.

\bibitem[Zhang et~al.(2023)Zhang, Xu, Fu, Wu, Liu, Gao, Zhao, Wan, Wan, Lu, et~al.]{zhang2023toward}
Shaoqing Zhang, Shiming Xu, Haohuan Fu, Lixin Wu, Zhao Liu, Yang Gao, Chun Zhao, Wubing Wan, Lingfeng Wan, Haitian Lu, et~al.
\newblock Toward earth system modeling with resolved clouds and ocean submesoscales on heterogeneous many-core hpcs.
\newblock \emph{National Science Review}, page nwad069, 2023.

\bibitem[Satoh et~al.(2008)Satoh, Matsuno, Tomita, and et. al.]{satoh2008nonhydrostatic}
Masaki Satoh, Taro Matsuno, Hirofumi Tomita, and et. al.
\newblock {Nonhydrostatic icosahedral atmospheric model (NICAM) for global cloud resolving simulations}.
\newblock \emph{Journal of Computational Physics}, 227\penalty0 (7):\penalty0 3486--3514, 2008.

\bibitem[Miyamoto et~al.(2013)Miyamoto, Kajikawa, Yoshida, Yamaura, Yashiro, and Tomita]{miyamoto2013deep}
Yoshiaki Miyamoto, Yoshiyuki Kajikawa, Ryuji Yoshida, Tsuyoshi Yamaura, Hisashi Yashiro, and Hirofumi Tomita.
\newblock Deep moist atmospheric convection in a subkilometer global simulation.
\newblock \emph{Geophysical Research Letters}, 40\penalty0 (18):\penalty0 4922--4926, 2013.

\bibitem[Dennis et~al.(2012)Dennis, Edwards, Evans, Guba, Lauritzen, Mirin, St-Cyr, Taylor, and Worley]{dennis2012cam}
John~M Dennis, Jim Edwards, Katherine~J Evans, Oksana Guba, Peter~H Lauritzen, Arthur~A Mirin, Amik St-Cyr, Mark~A Taylor, and Patrick~H Worley.
\newblock Cam-se: A scalable spectral element dynamical core for the community atmosphere model.
\newblock \emph{The International Journal of High Performance Computing Applications}, 26\penalty0 (1):\penalty0 74--89, 2012.

\bibitem[Johnsen et~al.(2013)Johnsen, Straka, Shapiro, Norton, and Galarneau]{johnsen2013petascale}
Peter Johnsen, Mark Straka, Melvyn Shapiro, Alan Norton, and Thomas Galarneau.
\newblock {Petascale WRF simulation of hurricane sandy: Deployment of NCSA's cray XE6 blue waters}.
\newblock In \emph{High Performance Computing, Networking, Storage and Analysis (SC), 2013 International Conference for}, pages 1--7. IEEE, 2013.

\bibitem[Golaz et~al.(2022)Golaz, Van~Roekel, Zheng, Roberts, Wolfe, Lin, Bradley, Tang, Maltrud, Forsyth, et~al.]{golaz2022doe-e3sm2}
Jean-Christophe Golaz, Luke~P Van~Roekel, Xue Zheng, Andrew~F Roberts, Jonathan~D Wolfe, Wuyin Lin, Andrew~M Bradley, Qi~Tang, Mathew~E Maltrud, Ryan~M Forsyth, et~al.
\newblock The doe e3sm model version 2: overview of the physical model and initial model evaluation.
\newblock \emph{Journal of Advances in Modeling Earth Systems}, 14\penalty0 (12), 2022.

\bibitem[Bertagna et~al.(2020)Bertagna, Guba, Taylor, Foucar, Larkin, Bradley, Rajamanickam, and Salinger]{bertagna2020performance}
Luca Bertagna, Oksana Guba, Mark~A Taylor, James~G Foucar, Jeff Larkin, Andrew~M Bradley, Sivasankaran Rajamanickam, and Andrew~G Salinger.
\newblock A performance-portable nonhydrostatic atmospheric dycore for the energy exascale earth system model running at cloud-resolving resolutions.
\newblock In \emph{SC20: International Conference for High Performance Computing, Networking, Storage and Analysis}, pages 1--14. IEEE, 2020.

\bibitem[Edwards et~al.(2014)Edwards, Trott, and Sunderland]{edwards2014kokkos}
H~Carter Edwards, Christian~R Trott, and Daniel Sunderland.
\newblock Kokkos: Enabling manycore performance portability through polymorphic memory access patterns.
\newblock \emph{Journal of parallel and distributed computing}, 74\penalty0 (12):\penalty0 3202--3216, 2014.

\bibitem[Fuhrer et~al.(2018)Fuhrer, Chadha, Hoefler, Kwasniewski, Lapillonne, Leutwyler, L{\"u}thi, Osuna, Sch{\"a}r, Schulthess, et~al.]{fuhrer2018near}
Oliver Fuhrer, Tarun Chadha, Torsten Hoefler, Grzegorz Kwasniewski, Xavier Lapillonne, David Leutwyler, Daniel L{\"u}thi, Carlos Osuna, Christoph Sch{\"a}r, Thomas~C Schulthess, et~al.
\newblock Near-global climate simulation at 1 km resolution: establishing a performance baseline on 4888 gpus with cosmo 5.0.
\newblock \emph{Geoscientific Model Development}, 11\penalty0 (4):\penalty0 1665--1681, 2018.

\bibitem[Gysi et~al.(2015)Gysi, Osuna, Fuhrer, Bianco, and Schulthess]{gysi2015stella}
Tobias Gysi, Carlos Osuna, Oliver Fuhrer, Mauro Bianco, and Thomas~C Schulthess.
\newblock Stella: A domain-specific tool for structured grid methods in weather and climate models.
\newblock In \emph{Proceedings of the international conference for high performance computing, networking, storage and analysis}, pages 1--12, 2015.

\bibitem[Satoh et~al.(2017)Satoh, Tomita, Yashiro, Kajikawa, Miyamoto, Yamaura, Miyakawa, Nakano, Kodama, Noda, et~al.]{satoh2017outcomes}
Masaki Satoh, Hirofumi Tomita, Hisashi Yashiro, Yoshiyuki Kajikawa, Yoshiaki Miyamoto, Tsuyoshi Yamaura, Tomoki Miyakawa, Masuo Nakano, Chihiro Kodama, Akira~T Noda, et~al.
\newblock Outcomes and challenges of global high-resolution non-hydrostatic atmospheric simulations using the k computer.
\newblock \emph{Progress in Earth and Planetary Science}, 4\penalty0 (1):\penalty0 1--24, 2017.

\bibitem[Sato et~al.(2020)Sato, Ishikawa, Tomita, Kodama, Odajima, Tsuji, Yashiro, Aoki, Shida, Miyoshi, et~al.]{sato2020co}
Mitsuhisa Sato, Yutaka Ishikawa, Hirofumi Tomita, Yuetsu Kodama, Tetsuya Odajima, Miwako Tsuji, Hisashi Yashiro, Masaki Aoki, Naoyuki Shida, Ikuo Miyoshi, et~al.
\newblock Co-design for a64fx manycore processor and” fugaku”.
\newblock In \emph{SC20: International Conference for High Performance Computing, Networking, Storage and Analysis}, pages 1--15. IEEE, 2020.

\bibitem[Hohenegger et~al.(2023)Hohenegger, Korn, Linardakis, Redler, Schnur, Adamidis, Bao, Bastin, Behravesh, Bergemann, et~al.]{hohenegger2023icon-sapphire}
Cathy Hohenegger, Peter Korn, Leonidas Linardakis, Ren{\'e} Redler, Reiner Schnur, Panagiotis Adamidis, Jiawei Bao, Swantje Bastin, Milad Behravesh, Martin Bergemann, et~al.
\newblock Icon-sapphire: Simulating the components of the earth system and their interactions at kilometer and subkilometer scales.
\newblock \emph{Geoscientific Model Development}, 16\penalty0 (2):\penalty0 779--811, 2023.

\bibitem[Jungclaus et~al.(2022)Jungclaus, Lorenz, Schmidt, Brovkin, Br{\"u}ggemann, Chegini, Cr{\"u}ger, De-Vrese, Gayler, Giorgetta, et~al.]{jungclaus2022icon}
Johann~H Jungclaus, Stephan~J Lorenz, Hauke Schmidt, Victor Brovkin, Nils Br{\"u}ggemann, Fatemeh Chegini, Traute Cr{\"u}ger, Philipp De-Vrese, Veronika Gayler, Marco~A Giorgetta, et~al.
\newblock The icon earth system model version 1.0.
\newblock \emph{Journal of Advances in Modeling Earth Systems}, 14\penalty0 (4):\penalty0 e2021MS002813, 2022.

\bibitem[Z{\"a}ngl et~al.(2015)Z{\"a}ngl, Reinert, R{\'\i}podas, and Baldauf]{zangl2015icon-dycore}
G{\"u}nther Z{\"a}ngl, Daniel Reinert, Pilar R{\'\i}podas, and Michael Baldauf.
\newblock The icon (icosahedral non-hydrostatic) modelling framework of dwd and mpi-m: Description of the non-hydrostatic dynamical core.
\newblock \emph{Quarterly Journal of the Royal Meteorological Society}, 141\penalty0 (687):\penalty0 563--579, 2015.

\bibitem[Giorgetta et~al.(2022)Giorgetta, Sawyer, Lapillonne, Adamidis, Alexeev, Cl{\'e}ment, Dietlicher, Engels, Esch, Franke, et~al.]{giorgetta2022icon-gpu}
Marco~A Giorgetta, William Sawyer, Xavier Lapillonne, Panagiotis Adamidis, Dmitry Alexeev, Valentin Cl{\'e}ment, Remo Dietlicher, Jan~Frederik Engels, Monika Esch, Henning Franke, et~al.
\newblock The icon-a model for direct qbo simulations on gpus.
\newblock \emph{Geoscientific Model Development}, 15\penalty0 (18):\penalty0 6985--7016, 2022.

\bibitem[Wedi et~al.(2020)Wedi, Polichtchouk, Dueben, Anantharaj, Bauer, Boussetta, Browne, Deconinck, Gaudin, Hadade, et~al.]{wedi2020baseline}
Nils~P Wedi, Inna Polichtchouk, Peter Dueben, Valentine~G Anantharaj, Peter Bauer, Souhail Boussetta, Philip Browne, Willem Deconinck, Wayne Gaudin, Ioan Hadade, et~al.
\newblock A baseline for global weather and climate simulations at 1 km resolution.
\newblock \emph{Journal of Advances in Modeling Earth Systems}, 12\penalty0 (11):\penalty0 e2020MS002192, 2020.

\bibitem[Freitas et~al.(2020)Freitas, Putman, Arnold, Adams, and Grell]{Freitas2020GRL}
Saulo~R. Freitas, William~M. Putman, Nathan~P. Arnold, David~K. Adams, and Georg~A. Grell.
\newblock Cascading toward a kilometer-scale gcm: Impacts of a scale-aware convection parameterization in the goddard earth observing system gcm.
\newblock \emph{Geophysical Research Letters}, 47\penalty0 (17):\penalty0 e2020GL087682, 2020.
\newblock \doi{10.1029/2020GL087682}.

\bibitem[Fu et~al.(2016{\natexlab{b}})Fu, Liao, Yang, Wang, Song, Huang, Yang, Xue, Liu, Qiao, et~al.]{fu2016sunway}
Haohuan Fu, Junfeng Liao, Jinzhe Yang, Lanning Wang, Zhenya Song, Xiaomeng Huang, Chao Yang, Wei Xue, Fangfang Liu, Fangli Qiao, et~al.
\newblock The sunway taihulight supercomputer: system and applications.
\newblock \emph{Science China Information Sciences}, 59:\penalty0 1--16, 2016{\natexlab{b}}.

\bibitem[Zhang et~al.(2020)Zhang, Fu, Wu, Li, Wang, Zeng, Duan, Wan, Wang, Zhuang, et~al.]{zhang2020optimizing}
Shaoqing Zhang, Haohuan Fu, Lixin Wu, Yuxuan Li, Hong Wang, Yunhui Zeng, Xiaohui Duan, Wubing Wan, Li~Wang, Yuan Zhuang, et~al.
\newblock Optimizing high-resolution community earth system model on a heterogeneous many-core supercomputing platform.
\newblock \emph{Geoscientific Model Development}, 13\penalty0 (10):\penalty0 4809--4829, 2020.

\bibitem[Xu et~al.(2021)Xu, Ma, Zhou, Zhang, Liu, and Wang]{Xu2021GMD}
Shiming Xu, Jialiang Ma, Lu~Zhou, Yan Zhang, Jiping Liu, and Bin Wang.
\newblock Comparison of sea ice kinematics at different resolutions modeled with a grid hierarchy in the community earth system model (version 1.2.1).
\newblock \emph{Geoscientific Model Development}, 14\penalty0 (1):\penalty0 603--628, 2021.
\newblock \doi{10.5194/gmd-14-603-2021}.

\bibitem[Li et~al.(2021)Li, Duan, Gan, Wan, Chen, Xu, Yang, Liu, Xue, Fu, et~al.]{li2021enabling}
Yuxuan Li, Xiaohui Duan, Lin Gan, Wubing Wan, Yuhu Chen, Kai Xu, Jinzhe Yang, Weiguo Liu, Wei Xue, Haohuan Fu, et~al.
\newblock Enabling large-scale simulation of cam on the sunway taihulight supercomputer.
\newblock \emph{IEEE Transactions on Computers}, 71\penalty0 (4):\penalty0 824--837, 2021.

\bibitem[Waley et~al.(2005)]{waley2005analects}
Arthur Waley et~al.
\newblock \emph{The analects of Confucius}, volume~28.
\newblock Psychology Press, 2005.

\end{thebibliography}

\end{document}